\documentclass[twocolumn]{aastex61}
\pdfoutput=1 
\usepackage{amsmath,amstext,mathtools,graphicx,rotating}
\usepackage[T1]{fontenc}
\usepackage{apjfonts} 
\usepackage[figure,figure*]{hypcap}

\shorttitle{Inversions of complex M\lowercase{g} II \MakeLowercase{h}\&\MakeLowercase{k} profiles in flares}
\shortauthors{Sainz Dalda and De Pontieu}

\newcommand{\mgii}{Mg {\small II} h\&k}
\newcommand{\mguv}{Mg  {\small II} UV triplet}
\newcommand{\mguvtt}{Mg  {\small II} UV2\&3}
\newcommand{\mguvo}{Mg  {\small II} UV1}
\newcommand{\cii}{C {\small II} 1334 \& 1335 \AA}
\newcommand{\ciifour}{C {\small II} 1334 \AA}
\newcommand{\ciifive}{C {\small II} 1335 \AA}
\newcommand{\fei}{Fe {\small I}}
\newcommand{\feii}{Fe {\small II}}

\newcommand{\vturb}{$v_{turb}$}
\newcommand{\vlos}{$v_{los}$}
\newcommand{\nne}{$n_e$}
\newcommand{\ltau}{$log(\tau)$}
\newcommand{\kms}{$km~s^{-1}$}

\begin{document}

\title{Chromospheric thermodynamic conditions from inversions of complex M\MakeLowercase{g} II \MakeLowercase{h}\&\MakeLowercase{k} profiles observed in flares} 

\author[0000-0002-3234-3070]{Alberto Sainz Dalda}
\affil{Lockheed Martin Solar \& Astrophysics Laboratory, 3251 Hanover Street, Palo Alto, CA 94304, USA}
\affil{Bay Area Environmental Research Institute, NASA Research Park, Moffett Field, CA 94035, USA.}
\author[0000-0002-8370-952X]{Bart De Pontieu}
\affil{Lockheed Martin Solar \& Astrophysics Laboratory, 3251 Hanover Street, Palo Alto, CA 94304, USA}
\affil{Rosseland Center for Solar Physics, University of Oslo, P.O. Box 1029 Blindern, NO-0315 Oslo, Norway}
\affil{Institute of Theoretical Astrophysics, University of Oslo, P.O. Box 1029 Blindern, NO-0315 Oslo, Norway}

\correspondingauthor{Alberto Sainz Dalda}\email{sainzdalda@baeri.org\\asainz.solarphysics@gmail.com}


\begin{abstract}
The flare activity of the Sun has been studied for decades, using both space- and ground-based telescopes. The former have mainly focused on the corona, while the latter have mostly been used to investigate the conditions in the chromosphere and photosphere. The Interface Region Imaging Spectrograph (IRIS) instrument has served as a gateway between these two cases, given its capability to observe quasi-simultaneously the corona, the transition region, and the chromosphere using different spectral lines in the near- and far-ultraviolet ranges. IRIS thus provides unique diagnostics to investigate the thermodynamics of flares in the solar atmosphere.

In particular, the \mgii\ and \mguv\ lines provide key information about the thermodynamics of low to upper chromosphere, while the \cii\ lines cover the upper-chromosphere and low transition region. The Mg II h\&k and Mg II UV triplet lines show a peculiar, pointy shape before and during the flare activity. The physical interpretation, i.e., the physical conditions in the chromosphere, that can explain these profiles has remained elusive. Several numerical experiments with forward modeling or ad-hoc assumptions have partially succeeded in reproducing these profiles, although the derived thermodynamic parameters seem either unrealistic or difficult to interpret in terms of physical mechanisms.

In this paper, we show the results of a non-LTE inversion of such peculiar profiles. To better constrain the atmospheric conditions, the \mgii\ lines and the \mguv\ lines are simultaneously inverted with the \cii\ lines. This combined inversion leads to more accurate derived thermodynamic parameters, especially the temperature and the turbulent motions (micro-turbulence velocity).  We use an iterative process that looks for the best fit between the observed profile and a synthetic profile obtained by considering non-local thermodynamic equilibrium and partial frequency redistribution of the radiation due to scattered photons. This method is computationally rather expensive ($\approx 6~CPU-hour/profile$). Therefore, we use the k-means clustering technique to identify representative profiles and associated representative model atmospheres. By inverting the representative profiles with the most advanced  inversion code (STiC), we are able to conclude that these unique, pointy profiles are associated with a simultaneous increase of the temperature and the electron density in the chromosphere,  
while the micro-turbulence velocity has values between 5-15 km/s, which seem to be more realistic values than the ones suggested in previous work. More importantly, the line-of-sight velocity shows a large gradient along the optical depth in the high chromosphere. This seems to be the parameter that gives the pointy aspect to these profiles. 
\end{abstract}


\section{Introduction}
A flare is the release of magnetic energy as a consequence of reconnection in magnetically stressed coronal loops. The magnetic energy stored in these loops comes from the low solar atmosphere. Once it is released, it is  transferred to both the outer and the lower solar atmosphere in a variety of energy forms: radiation, thermal energy,  non-thermal energy (accelerated particles and turbulence), kinetic energy, large-scale Alfvén waves, and others. The most evident observational counterpart of this sudden release of energy is an enhancement in  radiation flux (intensity) in almost any spectral range. This description, although simplistic, allow us to picture the basic flare phenomena. Many complex physical processes however occur and need to be properly studied for a full understanding of this type of solar event. This complexity is also reflected in the observational data we have of flares.

\begin{figure}[t]
\begin{center}
\includegraphics[width=.48\textwidth]{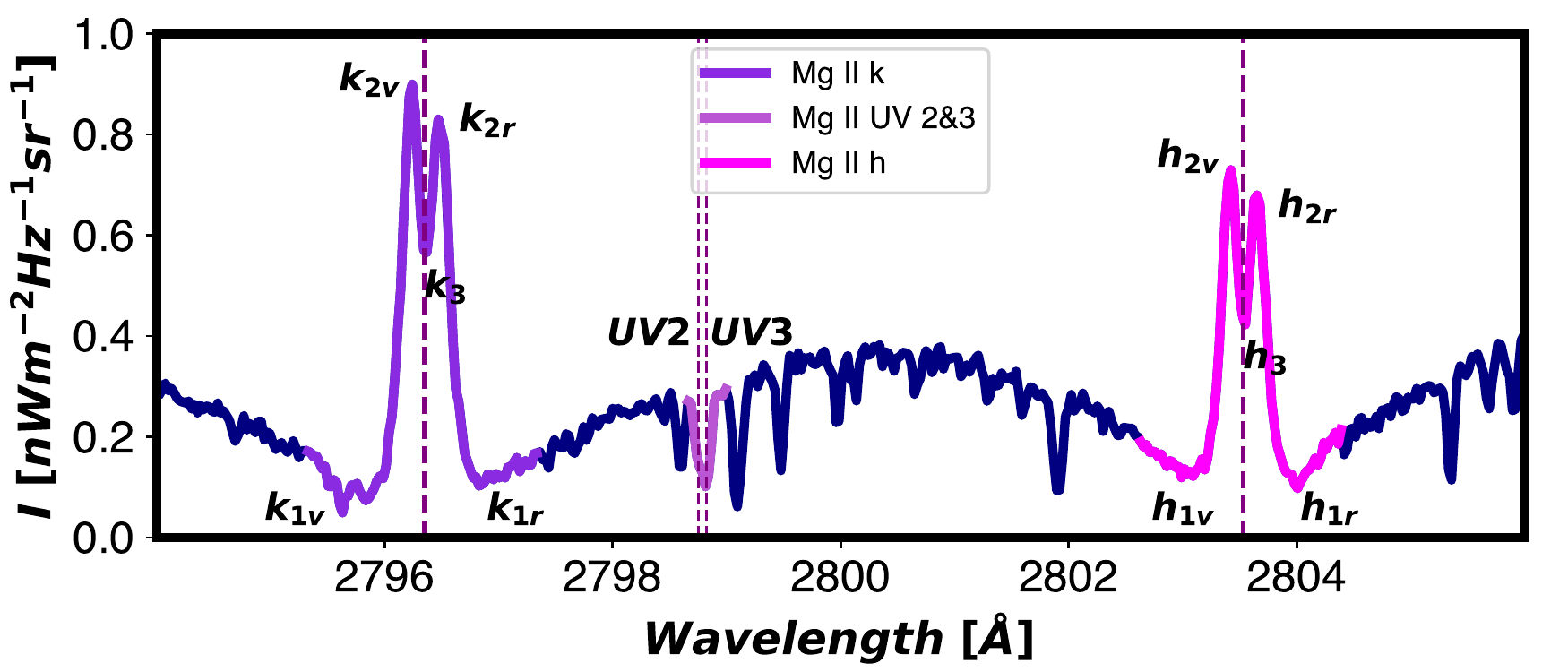}
\end{center}
\caption{ \mgii\ and \mguvtt\ lines  - the latter belonging  to the \mguv\ - as observed by IRIS in the quiet-sun. Spectral sampling is $0.025\AA$. The rest position for the core of the lines is indicated by the vertical dashed lines. The main features of the $h$ and $k$ lines are indicated with labels. The wavelength values are given in vacuum.}\label{fig:1}
\end{figure}

In the last few decades, with the advance of instrumentation, especially instrumentation onboard space-based observatories, we have largely improved our access to a steady flow of high-quality flare data. In this paper, we focus our attention on the interpretation of spectral signatures  in the \mgii\  and \cii\ lines observed by the Interface Region Imaging Spectrograph (IRIS, \cite{DePontieu14a}) during the X1.0-class flare of SOL2014-03-29T17:48. In particular, we will analyze the profiles in these lines corresponding to the flare ribbons duding the maximum of the flare. In this location, at that time, the \mgii\ profiles are characterized by a pointy, broad-on-the-base shape\footnote{By shape of a profile or a line, we mean the spectral distribution of the intensity (specifically, the spectral radiance) with respect to the wavelength in a given spectral range.}. Figure \ref{fig:1} shows a typical \mgii\ profile in the quiet Sun. The main features of the lines and their rest wavelength positions are indicated by labels and vertical lines respectively.

The  resonance \mgii\ profiles have been used in the past for the study of the chromosphere \citep{Lemaire73,Kohl76,Kneer81b,Lites82},  
including the study of flares \citep{Lemaire84}. The theoretical modeling and interpretation of these lines have been an active topic for decades \cite[e.g.,][]{Feldman77,Lites82,Lemaire83,Uitenbroek97}. More recently, thanks to the advance in the computational resources and triggered by the huge amount of data provided by IRIS, these lines have been investigated using more realistic  assumptions when solving the radiative transfer equation (RTE) \citep[e.g.,][]{Leenaarts13a,Leenaarts13b,Pereira13,Pereira15b,Sukhorukov17},  including treatment of polarized radiation \citep{delPinoAleman16,MansoSainz19}. 

In the {\it extreme} and {\it very pointy} profiles (see, e.g, Fig.~\ref{fig:typeA}) studied in this investigation, the depression (or self-reversal) in the core of the line ($k_{3}$ and $h_{3}$) has disappeared and the top of the profiles is defined by an inverted-V shape in just a few spectral samples. Note that some authors refer to  {\it single-peaked} \mgii\ profiles as those profiles that show $k_{3}$ and $h_{3}$ in emission, but at the same intensity  or slightly higher or lower than the  $k_{2}$ and $h_{2}$ spectral features. Such profiles, while being single-peaked, are mostly characterized by being {\it flat-topped}, as they were described by \cite{Carlsson15}. In contrast, the \mgii\ profiles discussed in this paper are {\it single-peaked} as well, but their main characteristics are:  i) their very pointy top, as a consequence of the  lack of $k_{3}$ and $h_{3}$ features and of having the violet and red components of the $k_{2}$ and $h_{2}$  almost totally blended in one pointy feature, ii) extended broad wings, which renders indistinguishable  the  $k_{1}$ and/or $h_{1}$ spectral features; and iii) the subordinated \mguv\  lines are in emission, in many cases also showing a pointy, broad shape, albeit not as extremely pointy, since the top of these lines shows an inverted-U shape.

The \cii\ lines also show a pointy shape, and in many cases it is red-shifted or showing a strong red-shifted component. \cite{Rathore15a} investigated the behavior of \cii\ lines in quiet Sun and found that the intensity profile may have both an optically thick and an optical thin component. These authors found, also through the analysis of the ratio of the intensity of the \cii\ lines,  that the single-peak profiles of \ciifive\ formed higher in the atmosphere than the single-peak profiles \ciifour.

These kind of pointy spectral line profiles, specially the \mgii, were identified in IRIS data as soon as the instrument started to observe 
flares.  \cite{Kerr15}  reported strong emission in the \mgii\ and the \mguvtt\ lines in the ribbons of the M class flare SOL2014-02-13T01:40 observed by IRIS. The authors noted the absence of the depression in the core of the \mgii\ lines, i.e., the lack of the $k_3$ and $h_3$ features. They also concluded, based on the ratio of the intensity between the $k$ and the $h$ lines, that these lines are optically thick during the flare. As we just mentioned above, this may not be the case for the \cii\ lines. The profiles shown by \cite{Kerr15} are inverted-U pointy profiles. Figure 6 of \cite{Liu15b} shows a selection of \mgii\ profiles belonging to the flare studied in the current paper. As we will discuss later, some of the profiles shown in that figure are  \textit{very pointy} profiles, while the ones shown in Figure 10 of that paper are \textit{extremely pointy} profiles or type A or B (see below).  \cite{Xu16} identified \textit{extremely pointy} \mgii\ and \cii\ profiles in the positive polarity part of the ribbon of the  M1.4-class  SOL2013-08-17T18:43 flare, but not in the negative polarity part (decrease in the contrast of the intensity). 

Several models have been proposed to explain the pointy profiles, particularly the ones observed by IRIS on SOL2014-03-29T17:48. \cite{RubiodaCosta16,RubiodaCosta17} studied the parameters needed to model the \mgii\ profiles observed during the maximum of this flare. The authors modified the thermodynamics parameters in hydrodynamic simulations (RADYN, \citealt{Carlsson95,Carlsson97a}), and then they obtained the synthetic profiles of these lines using the RH code \cite{Uitenbroek97} considering non-local thermodynamic equilibrium (non-LTE) and partial frequency redistribution of the scattered photons.  The authors obtained inverted-U, single \mgii\ profiles by increasing the temperature and density in the formation region of these lines. However, the intensity of these  profiles is larger than the intensity in the observed profiles.  They were only able to match intensity profiles with inverted-U, single peak profiles when they considered  a strong gradient in the line-of-sight velocity ($v_{los}$). These calculated profiles show however a significant asymmetry. In addition, their calculated profiles were not able to properly reproduce the large broad wings observed in the IRIS observations, except when they introduced micro-turbulent velocities ($v_{mic}$) values as large as $40~km~s^{-1}$, which they considered to be an unrealistic value. \cite{Zhu19}  followed a similar approach to that of \cite{RubiodaCosta16}, i.e., forward modeling using RADYN and RH, to interpret the \mgii\ lines of the same flare studied by \cite{RubiodaCosta16, RubiodaCosta17} and by us in the current paper. In this case, the authors considered the impact of the Stark effect on line broadening. They used the STARK-B database - in which line broadening is calculated based on a semi-classical impact-perturbation theory \cite{Dimitrijevic95,Dimitrijevic98}  - for the treatment of the quadratic Stark effect by RH. This is because the Stark effect previously implemented in RH considers the adiabatic approximation to calculate the quadratic Stark effect, and this approximation may underestimate the broadening for the \mgii\ lines during flares  \citep{RubiodaCosta17}. In summary,  \cite{Zhu19}  were able to fit the extremely pointy \mgii\ profiles observed by IRIS in the flare but only by considering an ad-hoc contribution of 30 times of the quadratic Stark effect obtained by using STARK-B in RH. This ad-hoc assumption allowed them to fit both the inverted-V, single peak, the intensity, and the broad wings. However, the authors were unable to provide a physical justification for the large ad-hoc enhancement of the Stark effect. They also tried to fit the broad wings by considering the STARK-B database value and various values of \vturb. They found an unrealistic value of \vturb $\approx 30~km~s^{-1}$ below the formation region of the line core of the \mgii\ lines, and even in this case, the fit in the part of the wings furthest from line core is not good both for \mgii\ and the \mguv.
\begin{figure*}[]
	\begin{center}
		\includegraphics[width=0.54\textwidth]{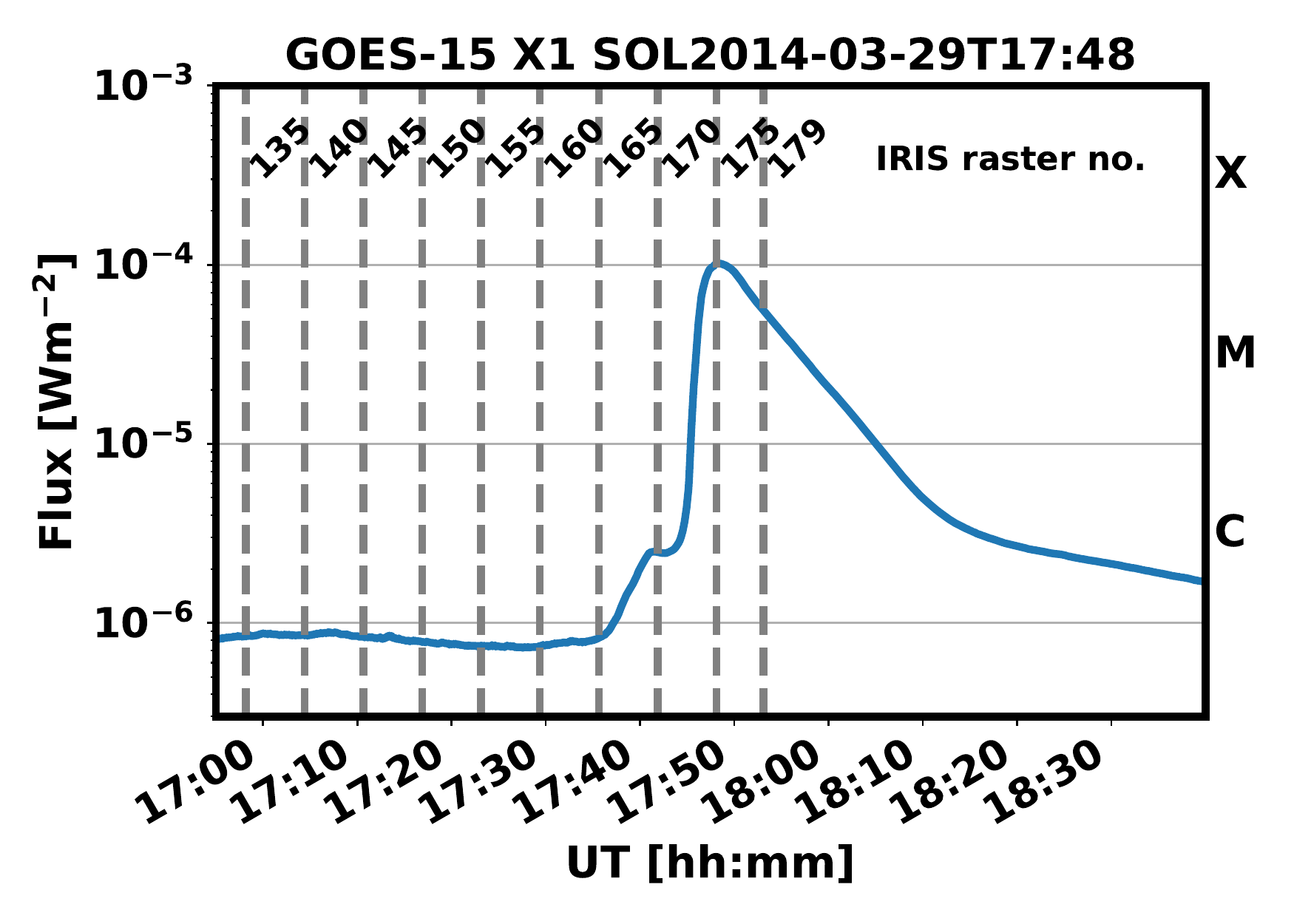}
		\includegraphics[width=0.45\textwidth]{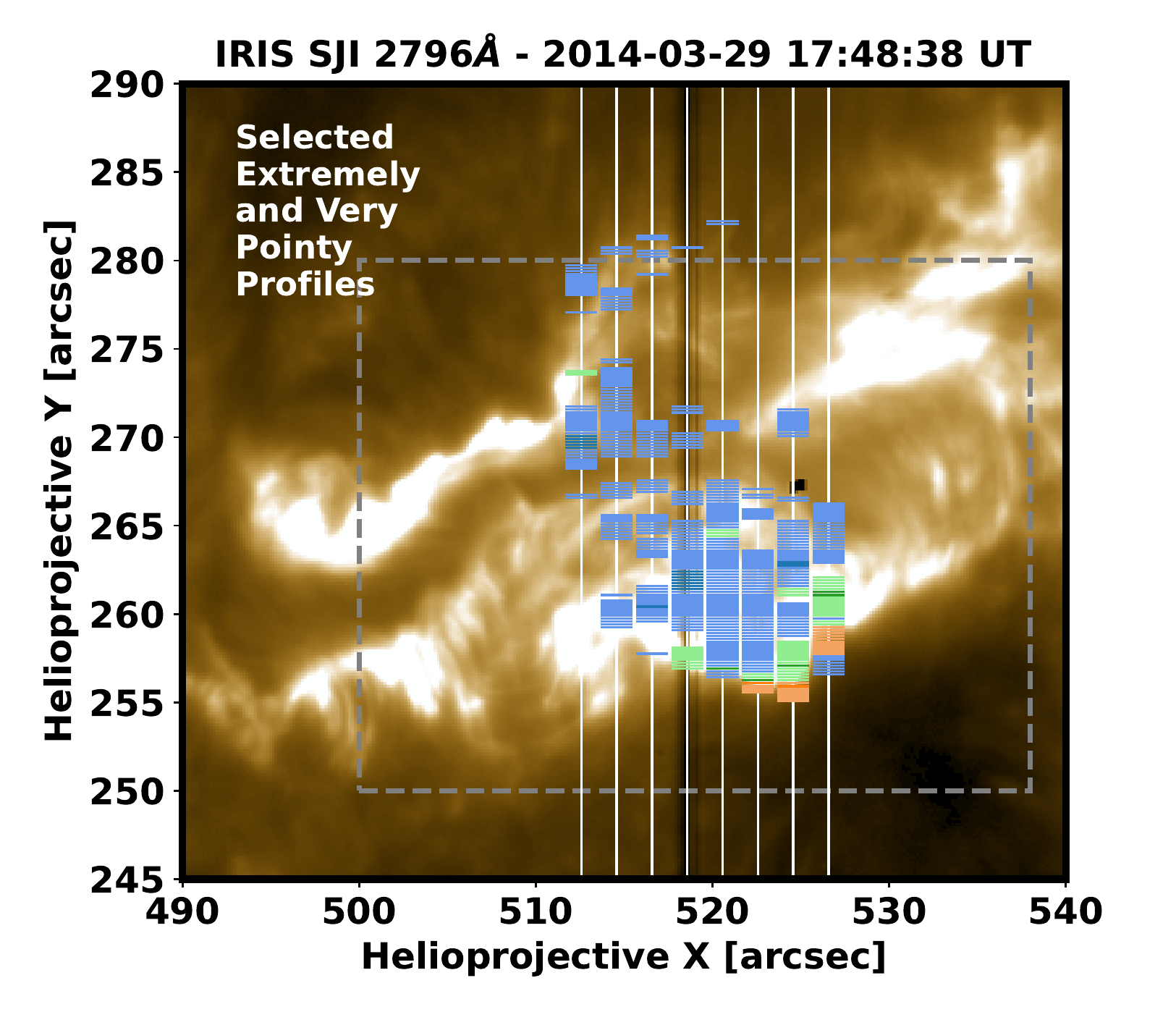}
	\end{center}
	\caption{Left: Temporal evolution of the X-ray flux during the X1-class flare at SOL2014-04-29T17:48. The vertical lines indicate the number of some IRIS rasters that recorded this flare.  Right: The image taken by the SJI instrument at IRIS during the maximum of the X1-class flare SOL2014-03-20T17:48 at $2796\AA$, that corresponds to the chromosphere. The time indicated at the top of the panel corresponds to the step number 4  of the raster. The location of the slits of all the steps of this raster (no. 175) are indicated with vertical lines. The colored squares mark the location of the extremely pointy profiles of the type A (blue) and the type B (orange), and of the very pointy or combined profiles (green). The locations in dark colors correspond to the profiles shown in figures \ref{fig:typeA}, \ref{fig:typeB} and \ref{fig:typeAB} respectively. The grey rectangle delimits the area shown in figure \ref{fig:allinv}.}\label{fig:goes_sji}
\end{figure*}
In this paper, we focus our attention on the study of both the extremely pointy and the very pointy profiles observed by IRIS during the maximum of the flare. This kind of profile is mostly present during the maximum of the flare. We note that while single-peaked profiles can also be seen in the pre-flare stage and they may help us to predict the flare onset with about 30-50  minutes advance notice \citep{Panos18, Woods21}, these are quite different: the peak of the latter \mgii\ profiles shows an inverted-U shape, the \mguv\ is barely in emission, and the \cii\ lines show a wide inverted-U shape.  The thermodynamics along the optical depth of these unique pre-flare profiles was for the first time  described by \cite{Woods21}.  In contrast, in the current paper we will reveal the thermodynamics of the most intriguing and peculiar profiles emitted at the maximum of the flare. 

The interpretation of these  extremely pointy profiles during the maximum of flares has presented a challenge for modelers and observers. In this paper, we  present a solution - in good agreement with several possibilities speculated in previous work - that is able to reproduce these complex profiles. This solution provides reasonable values of the thermodynamics parameters involved in the problem, i.e., gives a realistic view of the conditions in the chromosphere of flare during its maximum. 

\section{Materials and Methods}

The data analyzed in this paper correspond to the maximum of the X1-class flare at SOL2014-03-29T17:48. This data are part of the  multi-instrument observations that we led from the Dunn Solar Telescope at Sacramento Peak Observatory in coordination with IRIS and Hinode  \citep{Kosugi07}. Moreover, this flare was simultaneously observed by other observatories such as the  {\it Reuven Ramaty High-Energy Solar Spectroscopic Imager } (RHESSI, \citealt{Lin02}), the {\it Solar Dynamics Observatory} (SDO, \citealt{Pesnell12}), and the {\it Solar Terrestrial Relations Observatory} (STEREO, \citealt{Kaiser08}). A description of this unique observation and the evolution of the flare is provided in \cite{Kleint15}.  

The rasters obtained by IRIS of the  region where this flare ocurred - NOAA AR 12017 - span from 2014-03-29T14:09:39 UT to 2014-03-29T17:54:16 UT. This active region was located at $\mu=0.78$, with $\mu=cos\theta$, and $\theta$ the heliocentric viewing angle. The maximum X-ray flux measured by GOES-15 during the flare took place at 2014-03-29T17:48, which corresponds to raster no. 175 of the series of rasters taken by IRIS (see Figure \ref{fig:goes_sji}). Each raster consisted of an 8-step scan with the slit crossing the two ribbons of the flare. Each step of the raster is $2''$ in the direction perpendicular to the slit, covering a field-of-view (FoV) of $14'' \times 174''$, with $174''$ the length of the slit.  The exposure time was nominally $8s$ (but was reduced during the flare in response to the onboard automatic exposure control algorithm), the spectral sampling was $0.025\AA$ (in the NUV passband), and the spatial sampling along the slit was $0''.16$. At each step of the raster the IRIS Slit-jaw Imager (SJI) took an image. In this observation, during the acquisition at steps number 1, 5, and 7 the SJI took an image at  $1400\AA$, at step number 3 an image at $2832\AA$, and steps number 2, 4, 6 and 8 images at $2796\AA$. The FoV covered by the SJI was $167''\times174''$. The right panel in Figure \ref{fig:goes_sji} shows the SJI image taken by IRIS during the maximum of the flare at  $2796\AA$.

Since we are mostly interested in understanding, as accurately as possibly, the thermodynamics in the chromosphere, we decided to investigate simultaneously the \cii\ and the \mgii\ lines. It has been shown that these lines are sensitive to thermodynamics in roughly the same region of the chromosphere, both by solving the radiative transfer equation in 3D magnetohydrodynamics models  \citep{Rathore15a}, and by looking into the {\it mutual information} shared by these lines \citep{Panos21a}. Having this simultaneous information is a great advantage to decouple the  $T$ and \vturb\  encoded in the width of the spectral lines (see Section 4.6 in \citealt{Jefferies68}). 

In this paper we proceed in a similar fashion as \cite{Woods21} and invert simultaneously the \cii\ lines, the \mgii\ lines,  and the blended \mguvtt\ line. In addition, in this paper,  we invert the other line of the triplet, \mguvo. Thus, following the analysis made by \cite{Pereira15a} and \cite{Pereira15b}  on the formation region of the \mgii\ and the \mguv\  respectively, and by \cite{Rathore15a} on the \cii\ lines, we are in principle sampling the solar atmosphere from the low chromosphere to the top of the chromosphere. The caveat here is that some of the previous work focused on quiet Sun, and it is known that during flares the line formation can sometimes be significantly different.

\subsection{Data treatment: Clustering and Inversions}
To simultaneously invert these lines we have taken advantage of the capabilities of the {\it STockholm inversion Code} (STiC, \citealt{delaCruzRodriguez16,delaCruzRodriguez19}) to solve the radiative transfer equation for multiple lines and multiple atoms, considering non-local thermodynamic equilibrium and partial frequency redistribution of the radiation of the scattered photons. STiC uses the RH code \cite{Uitenbroek01} at the back-end to solve the RTE. STiC is the only code with the ability to properly treat the \mgii, \mguv, and  \cii\ lines to recover the thermodynamics encoded in these lines. However, to invert a single {\it joint profile} of these lines takes between 6 to 8 $CPU-hour$ - depending of the complexity of the profiles. Because of this computational burden, we have followed the strategy introduced by \cite{SainzDalda19}, i.e., to invert the {\it Representative Profile} (RP) and recover its corresponding {\it Representative Model Atmosphere} (RMA) by the inversion of the former. The RP is the averaged profile of a cluster of profiles that share the same shape, i.e. the same atmospheric conditions, since the shape of a profile is an encoded representation of the conditions of the matter and the radiation in the region where the lines are formed. To cluster the profiles we use the $k-means$ technique \citep{Steinhaus57,MacQueen67}. This technique, due to its simplicity and robustness, has become very popular in Machine Learning, and more recently in solar physics. However, clustering in solar data (Stokes V profiles) was already used in 2000 by \citep{SanchezAlmeida00}. Just like with any other method, care has to be taken when applying it to the data. It is important to understand how the method works. The core of the code is to find the centroids of clusters of elements so that the elements within a cluster are closer to its centroid than to any other centroid. For that, the code calculates the Euclidean distance (in the original $k-means$ version) between an initial number of elements randomly selected (in the original $k-means$ version) and all the elements in the data set.  Then, all the closest elements to a centroid form a cluster. A new centroid is calculated as the average of all the elements of that cluster. And again, the distance between all the elements in the data set and the centroids is calculated, and new centroids are calculated as the average of the new cluster. This process is repeated until the total sum of squared distance between the within-cluster samples and their corresponding centroid is minimized, that is:

\begin{equation}
	 \underset{\{C_i\}_{1}^{K}}{arg~min} \sum_{i=1} ^{K} \sum_{x_j\in C_i}||x_j - \mu_{i}||^{2}_{2}
\end{equation}

with $\lVert \cdot \rVert_2$ the $\ell_{2}$ norm, $\{C_i\}_{1}^{K}$ is a set of $K$ clusters $C$, $x_j$ the $j^{th}$ sample belonging to the $i^{th}$ cluster $C_i$, and $\mu_i$ the average of the samples $x$ belonging to the cluster $C_i$, i.e., the {\it centroid}. One of the challenges of this method is how to determine the number of clusters $K$ that are needed to properly cluster the data set. Several methods have been proposed to minimize the impact of this choice on the resulting clustering. We have selected the {\it elbow method} \citep{Thorndike53} to determine this number. It has been shown that a number of clusters larger than 100 is typically enough to cluster properly, in general, for IRIS \mgii\ data sets (see Section 1.2 of $IRIS^2$ tutorial web page\footnote{\href{https://iris.lmsal.com/iris2/iris2\_chapter01.html\#limitations-of-iris2-inversions}{https://iris.lmsal.com/iris2/iris2\_chapter01.html\#limitations-of-iris2-inversions}.}). Based on the results of the elbow plot\footnote{\href{https://www.youtube.com/watch?v=55sT144u5ag}{https://www.youtube.com/watch?v=55sT144u5ag}}, we have decided to cluster our data set in 320 RPs. Thus, we optimize the representation of the data  taking into account our computational resources (320 CPU cores). Note that what we call RPs are the centroids of the clusters of our data. In our case, a sample of our data set is the {\it joint profile} created by concatenating the profile of the \cii\ lines with those of the \mguvo line, and of the \mgii\ lines (including the \mguvtt\ blended line).  In machine learning jargon, the {\it dimensionality} of a sample is the number of features in the sample. In our case, the features are the intensity (radiative flux)  at the sampled wavelengths. Therefore, the dimensionality of our data set is the number of sampled wavelengths, and the number of samples is the number of spatial pixels where these profiles were simultaneously recorded. The latter is, for our data,  the common, co-aligned area recorded by IRIS in its far-ultraviolet (FUV) and the near-utraviolet  (NUV) detectors. For the case studied here, the number of samples is $\approx$8,000 $px$ per raster. Hence, to invert that number of profiles with STiC would take at least $\approx 48,000\ CPU-hours$. Nowadays, this number is not too large in terms of computation time. It would however turn computationally expensive if we wanted to invert the 175 available rasters of this flare  ($\approx8.4\ MCPU-hour$). Fortunately we can reduce these numbers very significantly through the use of representative profiles. As mentioned, a number of 320 RPs per raster is enough to cluster this kind of data, and they can be inverted with a mid-size server ($\approx1050\ CPU-hour$ for the 175 rasters). 

Another challenge of the $k-means$  method is related with the metric used to cluster. The Euclidean distance is not the best metric when the elements or samples of the data set have a large dimensionality. In our case, the dimensionality is 624, corresponding to 116, 36, and 472 sampled wavelengths for the \cii, \mguvo, and \mgii\ lines respectively. A reduction of the dimension of the profiles, e.g. using {\it Principal Components Analysis}, before clustering may mitigate this problem. 
In the test we have run, the effect of reducing the dimensionality before clustering has a negligible impact. This is probably because, despite having a large dimension, the  values of the features of our profiles are restricted to a small range of values.That means, for a given wavelength, the values  of the intensity  are in a well-defined and rather restricted range. As a consequence, the samples of our date set are located in a more or less well-defined region of the hyper-space defined by all ranges of all the features. However, we did adopt another approach to reduce the dimensionality. We have cropped the data around the lines, considering a spectral range large enough to capture both large blue and red shifts, eliminating the spectral positions that we are not interested in - basically those in the photospheric bump between the $k$ and $h$ lines. Note that we do include the \mguvtt, and some wavelengths in the far wings. 

Another issue of working with Euclidean distance is the scale of the features. If some features have very large values with respect to others, the smaller values will have a small impact on or significance in the distance. One way to solve this situation is to normalize each feature, for instance between 0 and 1. This is what we did for the cropped, joint profiles. These are the profiles that we have clustered. In this way, we give equal weight to all the involved spectral lines. After the $k-means$ is run over this new data set of joint profiles, the elements of a cluster are identified, that is, they are labeled with the number corresponding to that cluster. Finally, the RP of that cluster is calculated as the average of the original, joint profiles within the cluster.

This would be the standard way to proceed. But it is not the procedure we followed. During the maximum of the flare, the ratio between the maximum intensity in the \mgii\ lines and the \cii\ lines varies from $\sim$100 in the pseudo-quiet-sun (the closest area to the quiet in the FoV of our data) to just $\sim$10 in the ribbons. A significant variation also exists between the ratio of the integrated intensity of the \mguv\ lines and the \mgii\, although not as large. This is not important for clustering the data, since, as we mentioned above, we scale all the features between 0 and 1, but it represents a problem for the inversion. STiC tries iteratively to minimize the Euclidean distance between a synthetic profile - resulting from the synthesis of a model atmosphere- and the observed profile, slightly modifying the model atmosphere at each iteration. Again, since we are using the Euclidean distance as metric, we will have a problem if the scales of the features are very different. But now, due to a coding practicality, we cannot scale the profiles individually. In this case, we use a set of weights at the sample wavelengths to scale all the profiles. These weighted profiles are then inverted. It is usual to weight the spectral range corresponding to a line with respect other line, or in the case of spectropolarimetric profiles to scale the Stokes Q, U, and V profiles with respect to I, e.g. $w_I:w_Q:w_U:w_V$=1:50:50:10. In our case, we weight the \cii\ lines, \mguv\ lines  with respect to \mgii\ lines. However, given the large variation in the ratio of the integrated intensity of \cii\ lines with respect to the \mgii\ lines, we decide to stratify the data in 8 percentiles, with the 8 parts dividing the number density distribution. The 3 or 4 first portions have many original profiles, representing the pseudo-quiet-sun (the closest region to the quiet-sun in the field of view). Then, the rest of the portions are associated with the flare ribbons. Each of these portions have their own weight ratio $w_{\mgii}:w_{CII}:w_{\mguv}$. Once the portions are defined, 40 RPs are calculated following the procedure explained above: creation of joint, concatenated profile, scaling between 0 to 1, k-means calculation, rebuild the k-means with the original data. Then, the 40 RPs of each portion are inverted considering their chosen weights. Note that in the first portions, since they are the most populated ones, the RPs are clustering a large number of profiles, while the rest of portions are clustering a significantly smaller number of profiles. This is beneficial to our investigation,  since we will have a better representation by the RPs of the original profiles in the flaring areas.
At this point, the data are ready to be inverted. Table \ref{table:inversion} shows the number of nodes for each cycle considered during the inversions. 

\begin{table}[]
	\begin{center}
		\begin{tabular}{ccccc}
			No. Cycle	&	1 & 2 &  3 & 4 \\
			\hline
			$T$  &  4&  7 &  9&  13 \\
			$v_{turb}$	&  2 & 4 & 8 & 13\\
			$v_{los}$	& 2 & 4 & 8 & 13 \\
		\end{tabular}
		\caption{Number of nodes at each cycle for the thermodynamics variables considered during the inversions.}\label{table:inversion}
	\end{center}
\end{table}

\subsection{Selected pointy profiles}
\begin{figure*}[]
\begin{center}
\includegraphics[width=\textwidth]{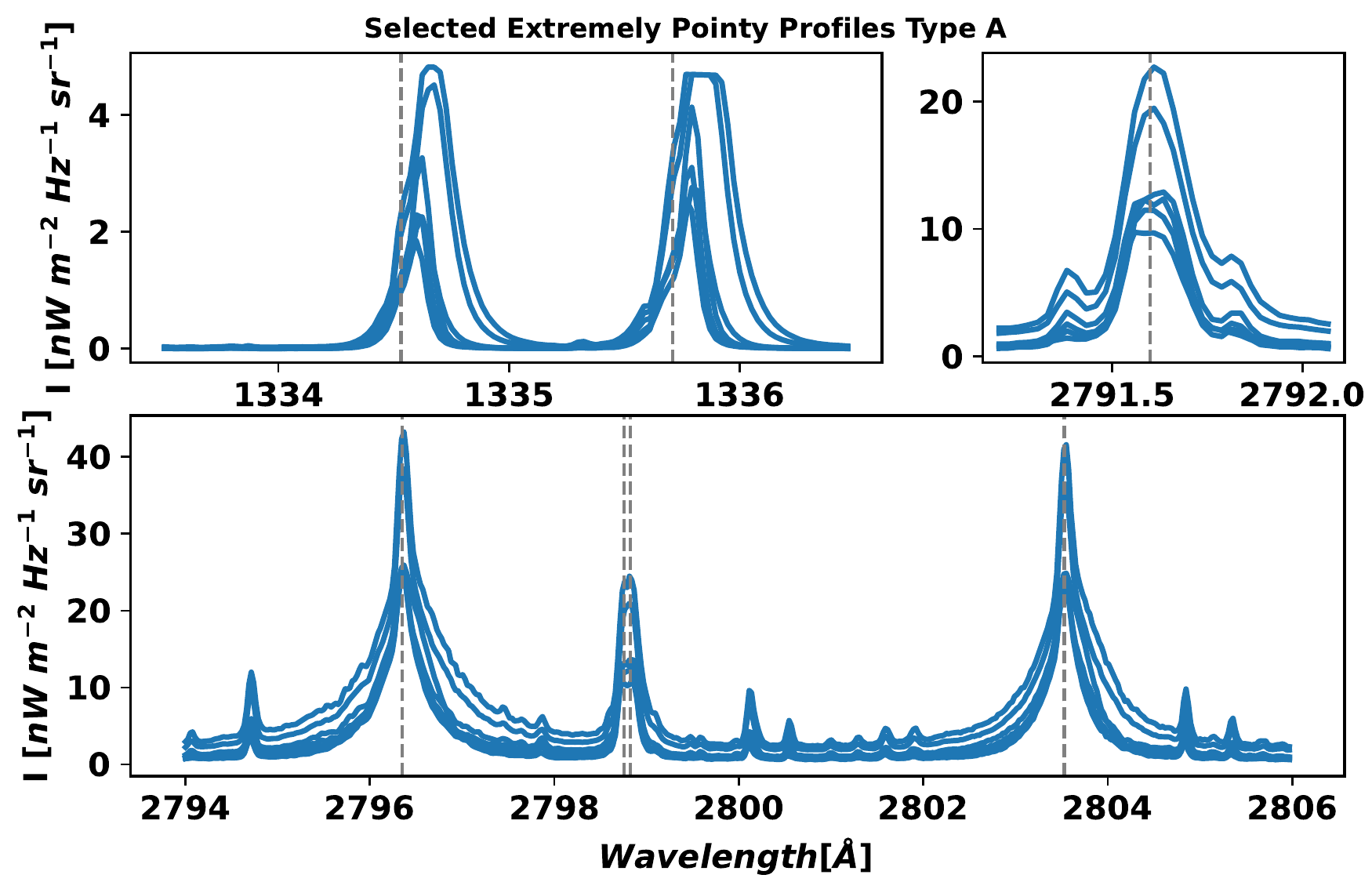}
\end{center}
\caption{Examples of the type A extremely pointy profiles.}\label{fig:typeA}
\end{figure*}

\begin{figure*}[]
	\begin{center}
		\includegraphics[width=\textwidth]{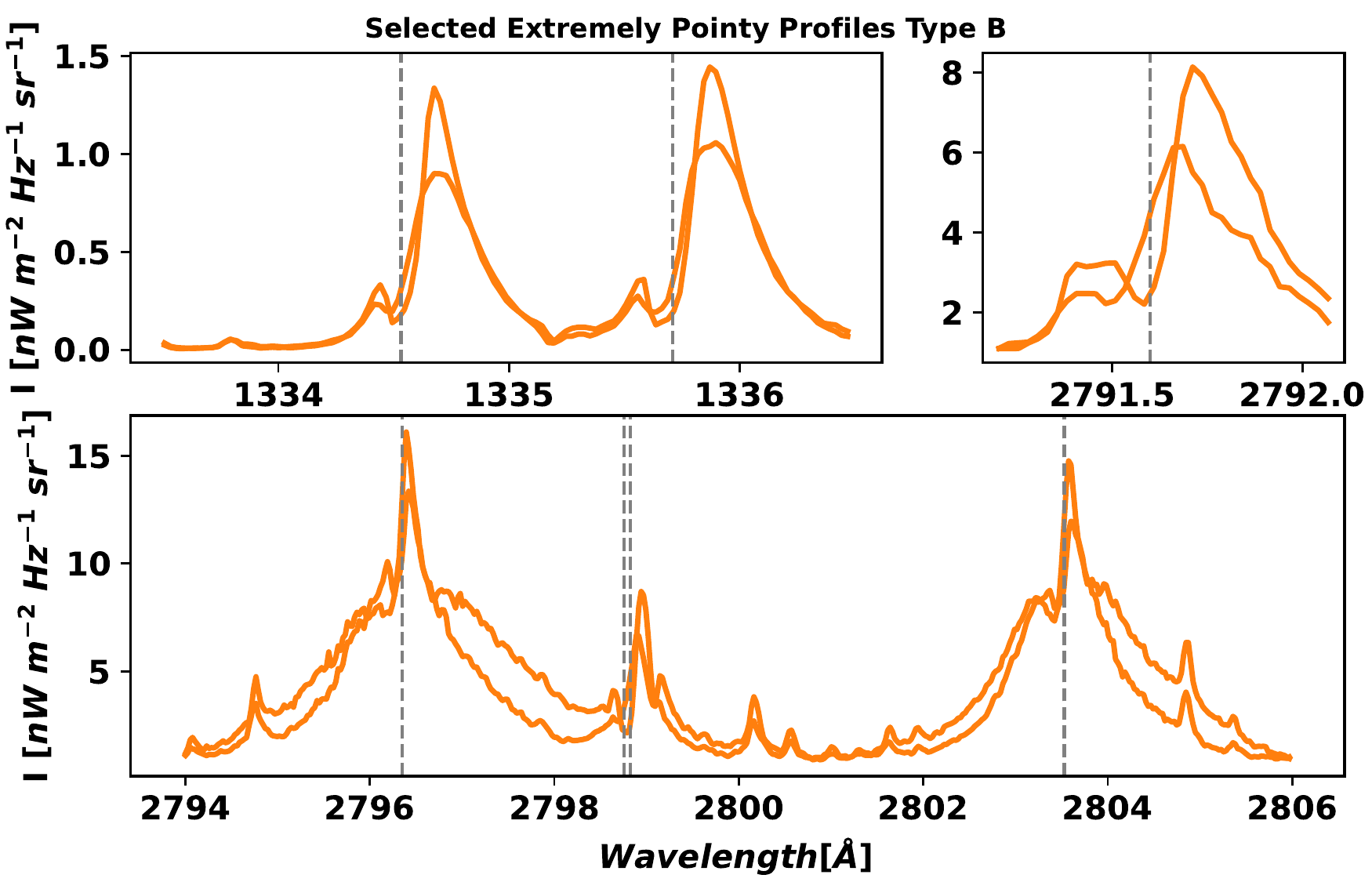}
	\end{center}
	\caption{Examples of the type B extremely pointy profiles.}\label{fig:typeB}
\end{figure*}

The selection of the "extremely" pointy profiles and the "very" pointy profiles was made through visual inspection. Thanks to the clustering of the data, this task is easily doable: we look for these peculiar profiles in a set of 320 RPs instead of $\approx$8,000 profiles. We have selected the most clear cases of these profiles to show their main characteristics. We can distinguish two main groups among the extremely pointy profiles that we have classified as Type A and Type B (see Figures \ref{fig:typeA} and \ref{fig:typeB}). The core of the \mgii\ extremely pointy profiles Type A resembles  the core of a Lorenztian distribution, while the wings look like the wings of a Laplacian distribution.  The \cii\ lines are pointy, red-shifted, and in most of the cases their shape shows a negative skew.  In some cases, the \cii\ lines saturate the response of the detector. 

The \mgii\ extremely pointy profiles type B (see Fig.~\ref{fig:typeB}), in addition to having  the pointy core, show very enhanced  wings.  Sometimes, the blue wing is slightly more enhanced than the red wing. The \mgii\ lines are a bit shifted to the red. The \mguv\ lines are clearly shifted to the red, and they have a blue component. Interestingly, the same happens for the \cii\ lines. 

\begin{figure*}[]
	\begin{center}
		\includegraphics[width=\textwidth]{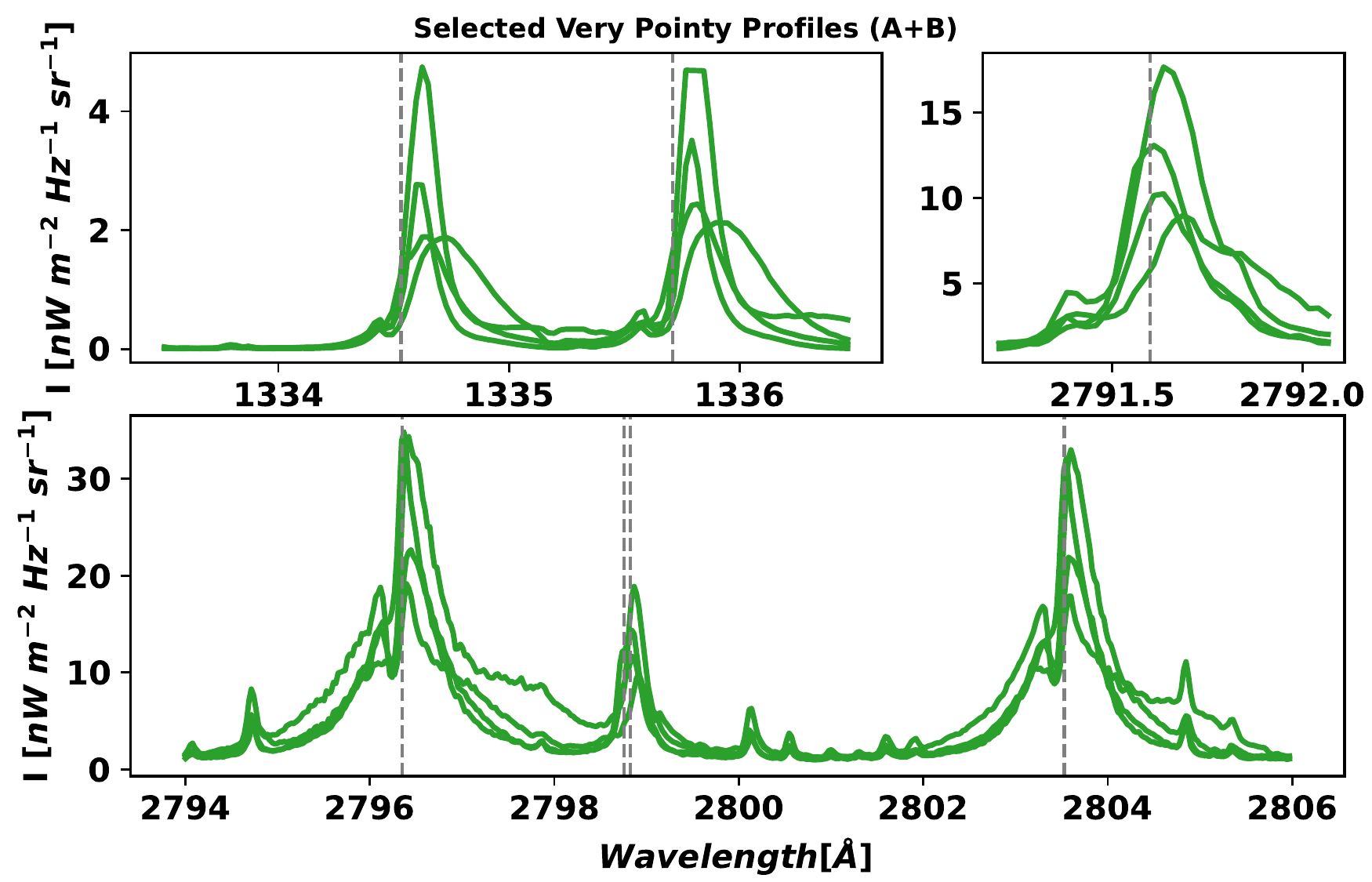}
	\end{center}
	\caption{Examples of the combined type of pointy profile.}\label{fig:typeAB}
\end{figure*}

The last type of profiles seems to be the {\it combination} of type A and B (see Fig.~\ref{fig:typeAB}). The main difference is the presence of a well defined blue component, and the peak of the line is now well-centered. Note that the blue component is well distinguished, and the profile can be clearly describe in terms of the $2v,2r,$ and $3$ features, that means in terms of 2 peaks and the central depression. In this case, the $2r$ largely dominates over the $2v$ feature, and it is located on the rest spectral position of the line. On the other hand, the $3$ feature is present, and it is slightly blue-shifted.

In all the types, the \mguv\ lines show either a blue or red component, or both (see Fig. \ref{fig:typeA}). These components appear, in most of the cases, both in the \mguvo\ line and the \mguvtt\ lines. This behavior indicates that these components are actually belonging to these lines and not to nearby lines in emission.

The spatial distribution of these profiles is shown in the right panel of Figure \ref{fig:goes_sji}. The Type A profiles are mostly located in the ribbon (dark blue ticks), the type B profiles are located in the leading edge (dark orange ticks), and the combination type profiles are mostly located close to (and just trailing) the leading edge (dark green ticks). The locations displayed in light colors show all the profiles of the various types not shown in Fig.~\ref{fig:typeA},~\ref{fig:typeB} and ~\ref{fig:typeAB}. The profiles associated with these locations have a range of gradually changing appearances within their corresponding type. For instance, a profile in a light blue location is an extremely pointy type A profile but the wings of the \mgii\ lines look more Lorentzian than Laplacian, in contrast to the most intense ones shown in Fig.~\ref{fig:typeA}, which have more Laplacian wings.

The pointy profiles almost always occur in the flare ribbons.
Pointy profiles have also been observed during the preflare phase, as we have already mentioned. However, it is during the maximum of the flare when the extreme and very pointy profiles appear, as right panel in Fig. \ref{fig:goes_sji} shows. 
	
\section{Results} 

\begin{figure*}[]
	\begin{center}
		\includegraphics[width=\textwidth]{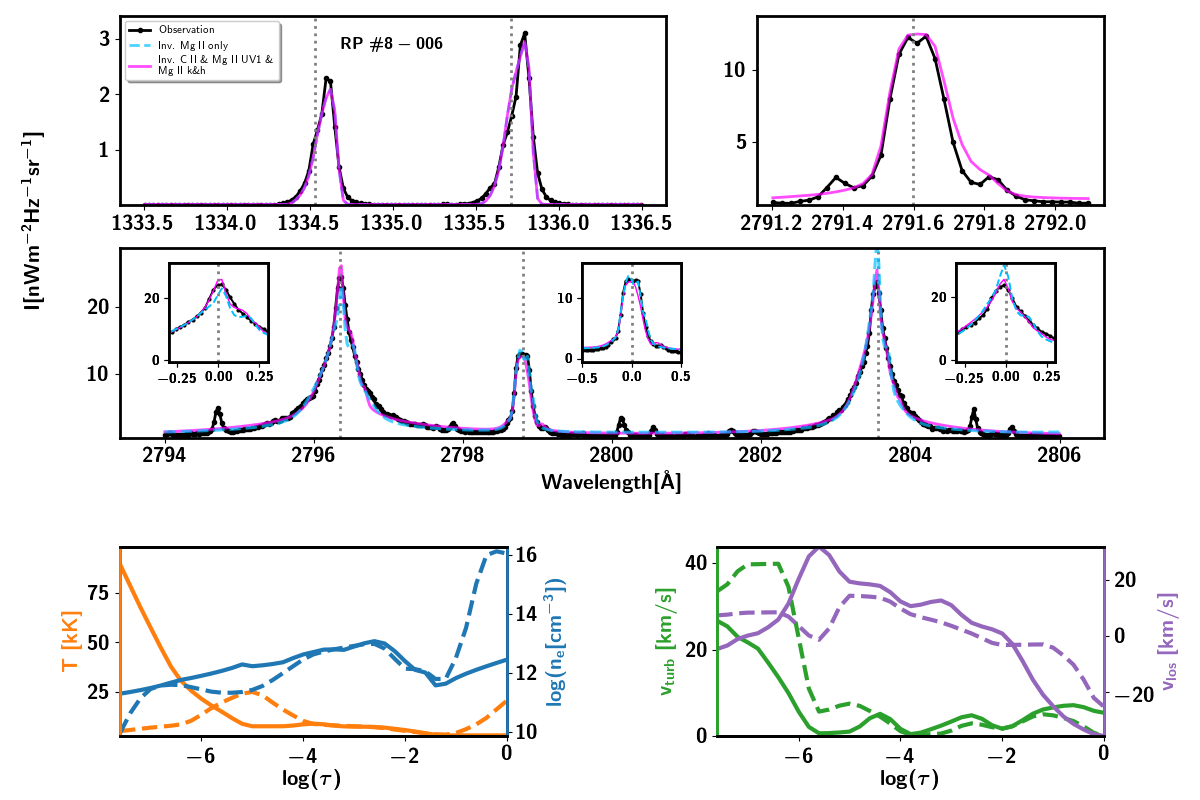}
	\end{center}
	\caption{Inversion (fit)  of the \cii, \mguvo\, \mgii, and \mguvtt\ lines of an extremely pointy profile type A, and the model recovered from the inversion. The three first panels show the inversion of an extremely pointy profile type A (in dotted, black line). The dashed, blue line corresponds to the inversion only taking into account the \mgii\ lines. The fuchsia line corresponds to the inversion considering simultaneously the \cii\ lines, the \mguvo\ line, and the \mgii\ lines - including the \mguvtt\ lines. The last two panels show the thermodynamic variables obtained from the inversions: temperature ($T$, in orange), logarithm of the electron density (\nne, in blue), velocity of turbulent motions or micro-turbulence ($v_{turb}$, in green), and line-of-sight velocity (\vlos, in violet). The dashed lines correspond to model recovered from the inversion considering only the \mgii\ lines, while the solid lines correspond to the inversion considering all the lines mentioned above.}\label{fig:inv_typeA_1}
\end{figure*}

\begin{figure*}[]
	\begin{center}
		\includegraphics[width=\textwidth]{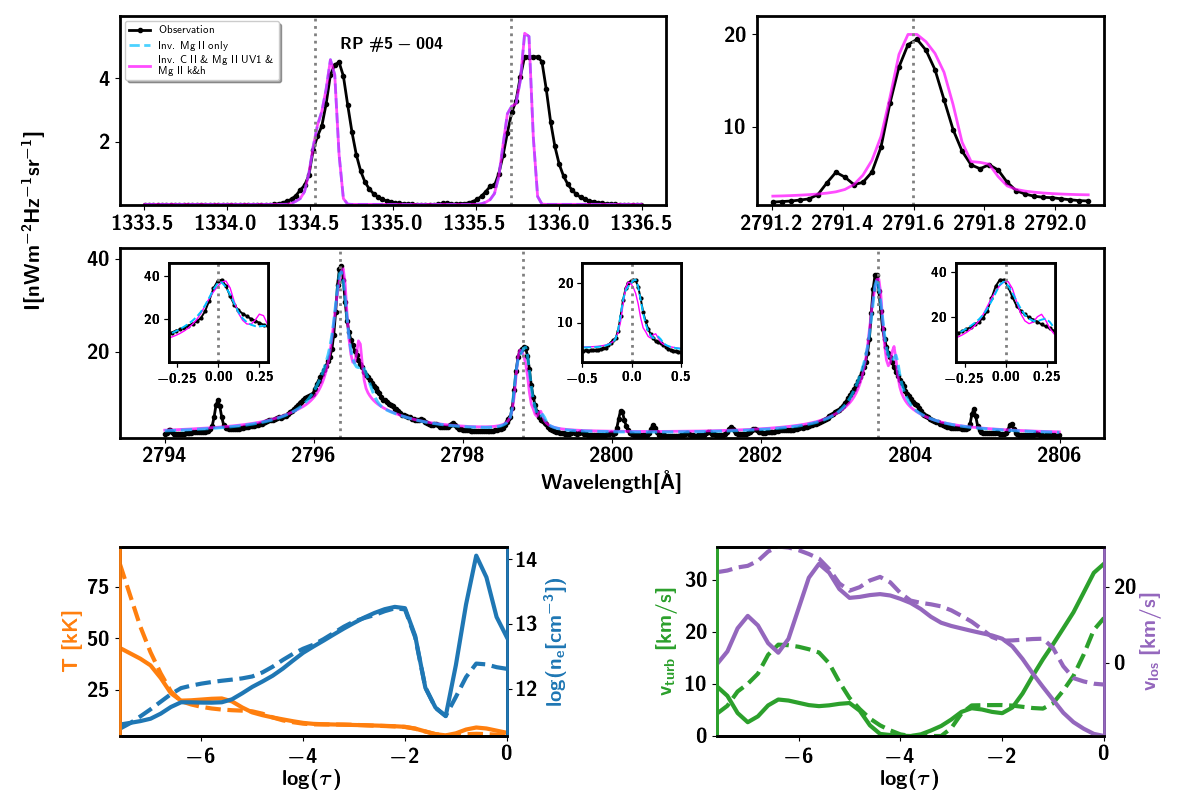}
	\end{center}
	\caption{Inversion (fit)  of the \cii, \mguvo\, \mgii, and \mguvtt\ lines of another extremely pointy profile type A, and the model recovered from the inversion. See caption of Figure \ref{fig:inv_typeA_1} for details.}\label{fig:inv_typeA_2}
\end{figure*}

\begin{figure*}[]
	\begin{center}
		\includegraphics[width=\textwidth]{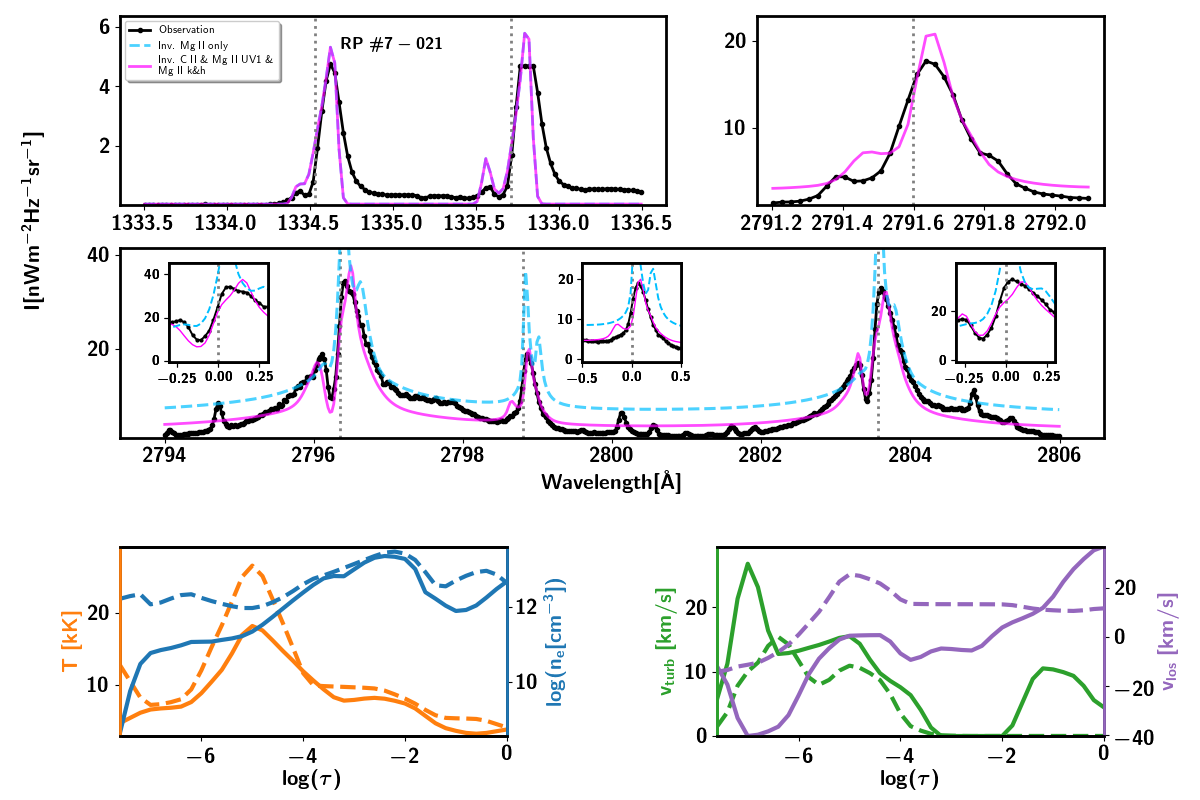}
	\end{center}
	\caption{Inversion (fit)  of the \cii, \mguvo\, \mgii, and \mguvtt\ lines of an extremely pointy profile type B, and the model recovered from the inversion. See caption of Figure \ref{fig:inv_typeA_1} for details.}\label{fig:inv_typeB}
\end{figure*}

\begin{figure*}[h!]
	\begin{center}
		\includegraphics[width=\textwidth]{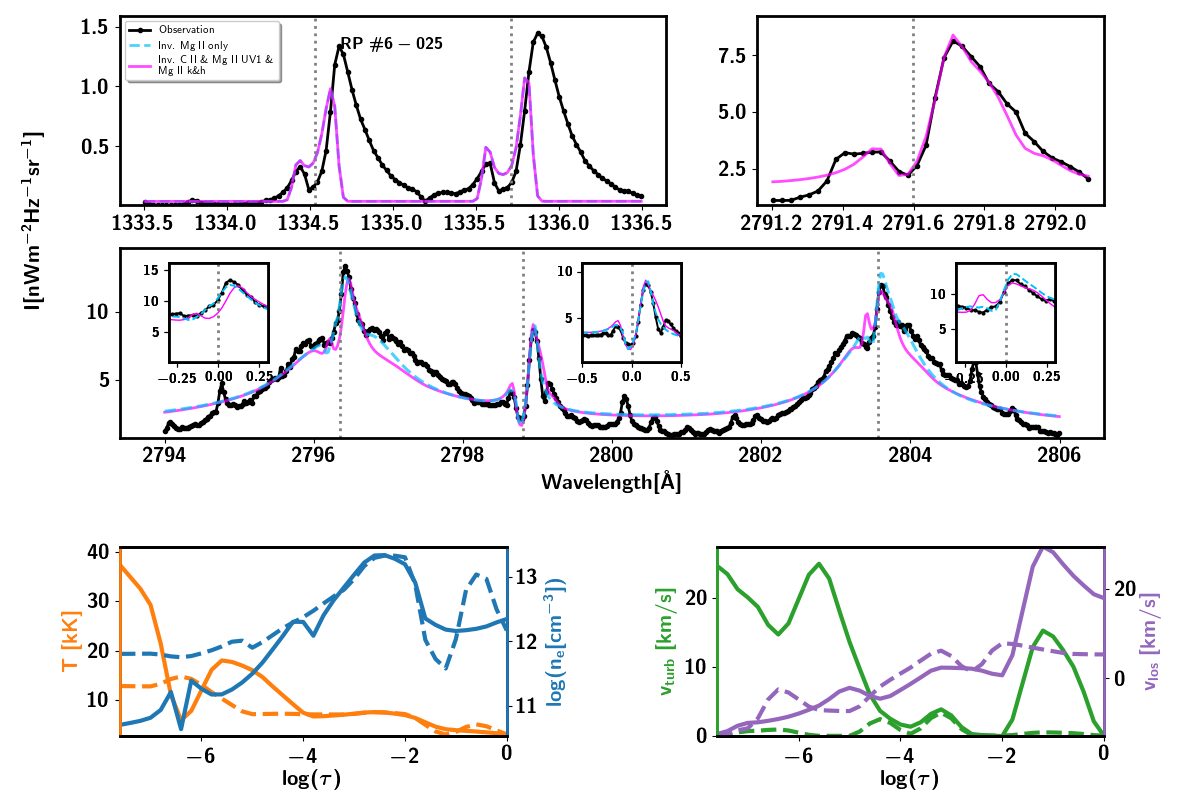}
	\end{center}
	\caption{Inversion (fit)  of the \cii, \mguvo\, \mgii, and \mguvtt\ lines of a combined pointy profile type, and the model recovered from the inversion.See caption of Figure \ref{fig:inv_typeA_1} for details.}\label{fig:inv_typeAB}
\end{figure*}

Figures \ref{fig:inv_typeA_1}, \ref{fig:inv_typeA_2}, \ref{fig:inv_typeB}, and \ref{fig:inv_typeAB} show the results from the inversions of two profiles of type A, and one of type B, and a combination profile, respectively. From top to bottom, from left to right, in each figure, the three first panels show the observed profile (in black dotted line)  and the inverted profile (fuchsia)
for the \cii\, the \mguvo, and the \mgii\ and \mguvtt\ lines respectively. In the second row, we also include the inversion for the case that only the \mgii\ and \mguvtt lines are inverted (dashed blue)\footnote{For the sake of simplicity, since the inversion of the Mg II h\&k lines and the Mg II UV2\&3 lines are always performed simultaneously, we will refer to this inversion as  Mg II h\&k-{\it only inversion}}. Thus, we can appreciate the effect of including the \cii\ lines (and the \mguvo line)  in the inversion to  recover the model atmosphere. The two last panels show the $T$ (left panel, left axis, in orange), the \nne (left panel, left axis, in blue), the \vturb (left panel, right axis, in green), and the \vlos (right panel, right axis, in violet). All these values are given with respect to the logarithm of the optical depth, $\log(\tau)$\footnote{In this paper, $\log(\tau)$ actually means $\log_{10}(\tau_{500})$, that means, the reference of the optical depth unity corresponds to the continuum at $500\ nm$.}. The dashed lines in these panels refer to the thermodynamic parameters corresponding to  the \mgii-only inversion. If the dashed blue line is not shown in the \mgii panel (third  panel), then the \mgii-onliy inversion failed for the $4^{th}$ inversion cycle and the model atmosphere shown in the bottom panels correspond to the $3^{rd}$ cycle.

The fit of the lines in Figure  \ref{fig:inv_typeA_1}, while not  perfect, is rather good. Both \cii\ lines are well fitted, with the inverted profile capturing the intensity, the shift, the skewness, and fitting the wings relatively well. The fit of the \mguvo 
is capturing the width of the line, and the intensity, although it is missing the little depression in the core and the little bumps present both in the blue and the red wing. The fit of the \mgii\ and the \mguvtt\ lines is very good for both the multi-line inversion and the \mgii-only inversion. In this case, the fit of the  multi-line inversion is better than the \mgii-only inversion. However, that is not always the case (see below). The $T$ shows very large values, between $7.5-20kK$ at
$-6.5 <$ \ltau $< -6.0$. Given the goodness of the fit both in the \cii\ lines and the \mgii\ lines, it is difficult to argue why such  high values would be incorrect. The \nne\ shows a steady increase from $-6.5<$ \ltau $<-2.5$.  The \vturb\ has values $\approx 20 km~s^{-1}$ in the high chromosphere\footnote{Roughly speaking, we define in this paper the high chromosphere as the optical depth range $-6.5<$ \ltau $< -5$, the mid chromosphere is $-5<$ \ltau $< -4$, the low chromosphere is  $-4<$ \ltau $< -2$, the high photosphere is $-2<$ \ltau $<-1$, and the low photosphere is $-1<$ \ltau $<0$. These ranges are dynamically changing, especially for events such as flares. Throughout such changes, the region just above the temperature minimum is referred to as the low chromosphere, while the steep, large increase of the temperature at small \ltau values is the top of the chrosmosphere.} (\ltau $=~-6$), and a wavy behavior at
$-6 <$ \ltau $<0$. The \vlos\ shows an important downflow gradient in the high chromosphere, from $0$ to $+20~$\kms. Then, from $-6 <$ \ltau $<-2$, the \vlos\ drops down to $0~$\kms\ in a slightly oscillatory way. At $-2<$ \ltau, there is a significant upflow gradient, from 0 to $-20~$\kms. Figure \ref{fig:inv_typeA_2} shows  a similar behavior in $T$ and \nne. The \vturb, in this case, peaks at \ltau = -5.6 with a value of $\approx 20~$\kms. The \vlos shows a steady downflow that 
increases from a $0~$\kms\  in the high chromosphere to $20~$\kms\ at \ltau=-1.8, followed by an upflow up to $-20~$\kms\ in the high-photosphere ($-2 <$ \ltau $< -1$). The optical depth where the flow turns from downflow to upflow in both cases is where the minimum of  $T$ and the decrease of \nne\ occurs. 

We note that we should not consider the values in the low-photosphere $-1 <$ \ltau $< 0$) as realistic, since the photospheric lines available in the IRIS \mgii\ spectral range do not show such a high Doppler shift.  We have tested that the photospheric flows ($-2 < $ \ltau) shown in the figures of this paper do not contribute to any aspects of the \cii\ lines and the \mgii\ lines. In some cases, e.g. Figure \ref{fig:inv_typeAB}, the velocity flows at $-2 <$ \ltau $< -1$ have an impact in the \mguv\ lines. That means, the velocities captured at this optical range are needed to explain the observed profiles, and therefore, despite their large values, considered as feasible. 

The values at \ltau $< -6.5$ should be taken with caution. This is because the current inversion scheme is not able to capture changes in the thermodynamics - if any - in that optical depth range from the \cii\ lines. None of the lines in this study are sensitive to changes in the thermodynamics at $-1 <$ \ltau. The values shown at these two ranges are due to the interpolation used by the inversion code in the nodes at that optical depths, and we have tested that changing those values in these ranges has no effect on the inverted profiles.



The profiles in Figure \ref{fig:inv_typeB} show an extremely pointy profile of type B. The $T$ peaks now a bit lower in the chromosphere, at \ltau$=-5$, but now the $T$ strikingly decreases from $-7<$\ltau$<-5$. The $T$ minimum and \nne\ depression is around \ltau $~=-1$.  Again, it is around that optical depth where the \vlos\ changes its sign. There is another significant difference, the flow is now the opposite: there is a decreasing upflow as the optical depth increase from the high chromosphere (\ltau $=-6.5$) towards the mid-low chromosphere ($-4 <$ \ltau $<-2$), where the \vlos\ is around $0~ $\kms. Then a downflow increases as the optical depth increases towards the photosphere. The turbulent motions in the high chromosphere have again a value $\approx -20~$\kms, being $\approx -10~$ \kms\ in the mid chromosphere and the photosphere. 

The model atmosphere associated with the profiles of the combined type  (see Figure \ref{fig:inv_typeAB}) is closer to the one associated to extremely profiles type B than to type A. Thus, the $T$ has a decrease again in the range $-6.5<$\ltau$<5.8$, then a peak in $\approx 20 kK$, with the minimum of $T$ and $n_e$ around 
\ltau$=-1.5$. The \vturb\ now peaks at \ltau$=-5.8$ with a value $\approx 25~$\kms, then decreases between 
$-4<$\ltau$<3$ to a value of $\approx 5~$\kms, and peaks at a value of $\approx 10~$\kms\ and $15~$\kms\ for the type B and the combined type respectively at the photosphere, which would explain the large broad wings of the \mgii\ lines. 

\subsection{How the extremely pointy profiles are formed}
\begin{figure*}[]
	\begin{center}
		\includegraphics[width=\textwidth]{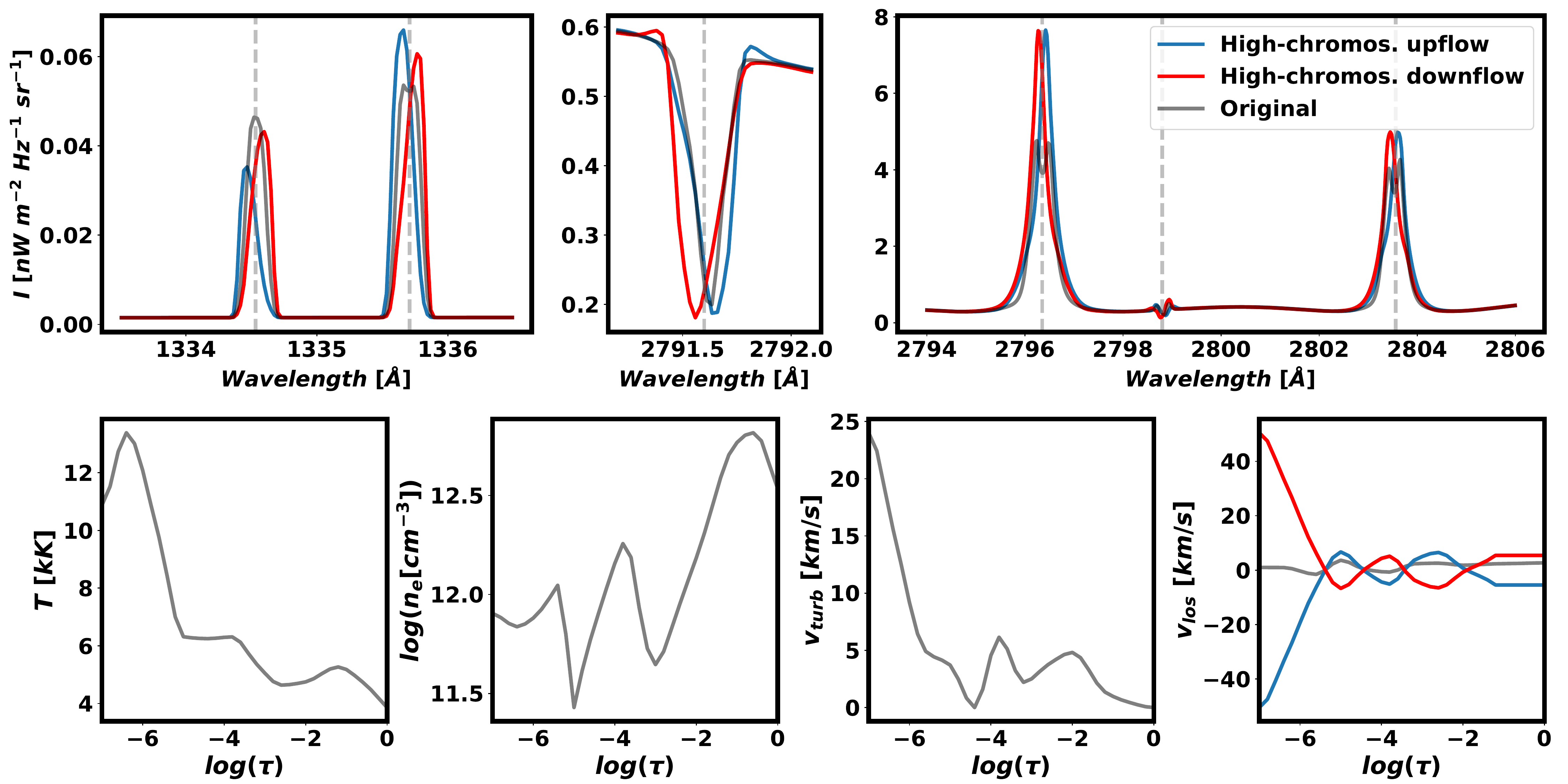}
	\caption{Creating an \mgii\ extremely pointy profile type A (red or blue) from a double-peaked \mgii\ profile (grey) by considering an extreme gradient downflow (red) or upflow (blue) in the high chromosphere.}\label{fig:building_pointyA}
	\end{center}
\end{figure*}

A simple way to create an extremely pointy type A profile in the \mgii\ lines is by considering a extreme gradient in the high chromosphere. Figure \ref{fig:building_pointyA} shows how to get an extremely pointy profile type A from a double-peaked \mgii\ profile. The red and blue profiles are the result of considering the $T$, \vturb, and \nne\ of the double-peaked profile (in grey) with the \vlos\ shown in red and blue lines respectively in the bottom, right-most panel of the Figure \ref{fig:building_pointyA}. In this example the steep gradient is located between $-7 <$ \ltau $< -5.5$ and goes from $~ \pm50$ to $0\  km~s^{-1}$. We have done several tests and we have also obtained extremely pointy profiles for lower values of \vlos.
Note that the \cii\ lines are shifted to the red and blue wavelengths for the downflow and upflow gradients respectively, and they show the skewness previously observed in the IRIS data. 

\begin{figure*}[]
	\begin{center}
		\includegraphics[width=\textwidth]{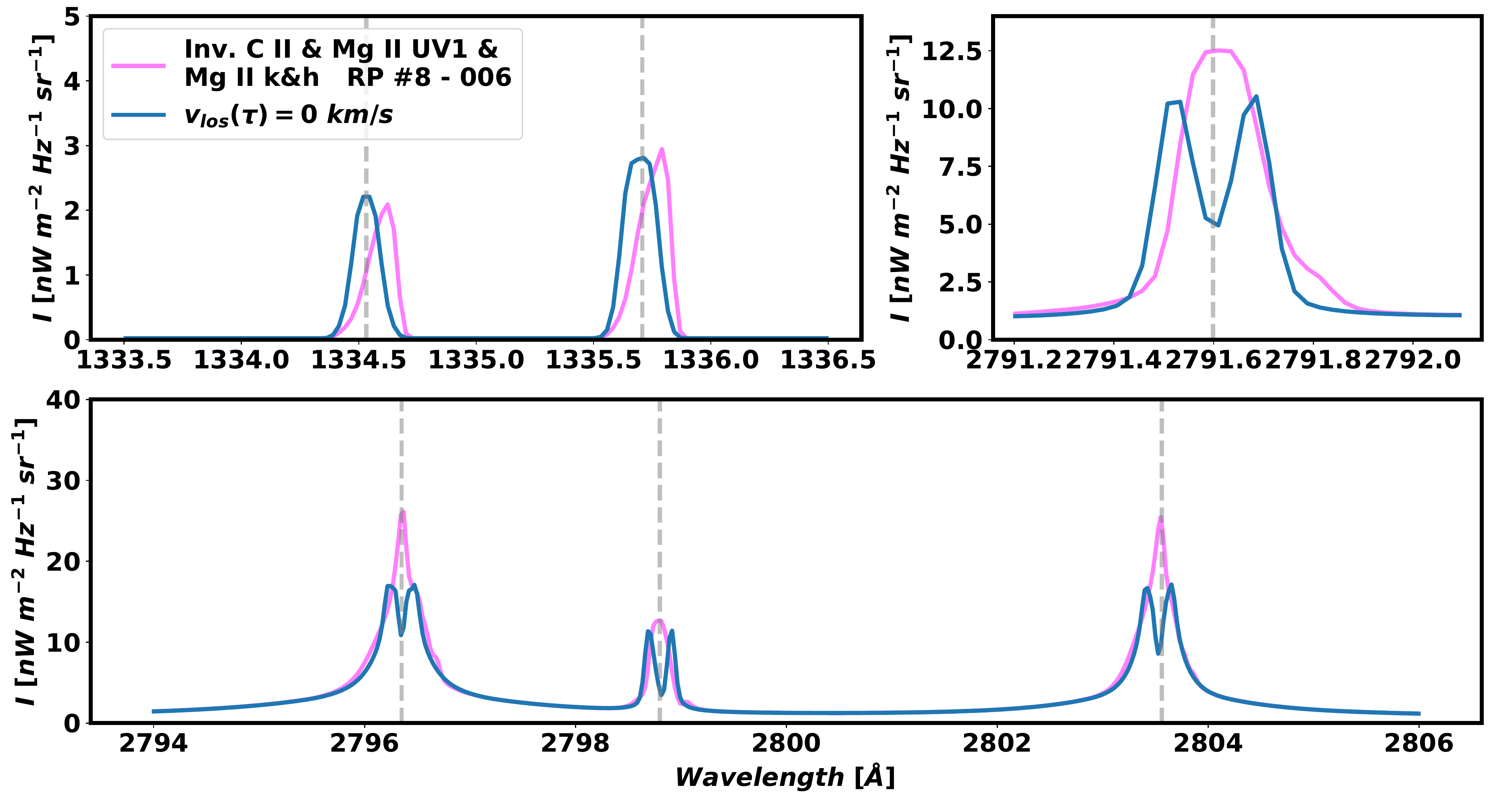}
	\caption{Deconstruction of the synthetic extremely pointy type A profile (in fuchsia) shown in the Figure \ref{fig:inv_typeA_1} that shows the role played by the \vlos. The profile in blues shows the resulting profile by considering \vlos $=0~km~s^{-1}$ in the atmosphere associated to the profile mentioned above (two bottom panels in the Figure \ref{fig:inv_typeA_1}).}.\label{fig:deconstruction_pointyA}
	\end{center}
\end{figure*}

Another way of studying what can cause pointy profiles is by studying the contribution of \vlos\ to the synthetic profile obtained by the inversion of the extremely pointy type A profile shown in Figure \ref{fig:inv_typeA_1} (in fuchsia). We have taken the same thermodynamics atmosphere associated with that profile (see bottom panels in the \ref{fig:inv_typeA_1}) and imposed a zero velocity at all optical depths ($v_{los}(\tau) = 0\ km~s^{-1})$, and then synthesized that new atmosphere. Again, we obtain a double-peaked profile shown in Figure \ref{fig:deconstruction_pointyA}.

In summary, the extremely pointy shape of these profiles is the signature of an extreme gradient in the  \vlos\ in the chromosphere. In the context of a flare, the downflows and upflows are associated with the {\it chromospheric condensation} and {\it evaporation} respectively. In this case, these flows have extreme gradients along the optical depth and take place in a well-determined optical depth range: the high chromosphere.

\section{Discussion}

\begin{figure*}[]
	\begin{center}
		\includegraphics[width=\textwidth]{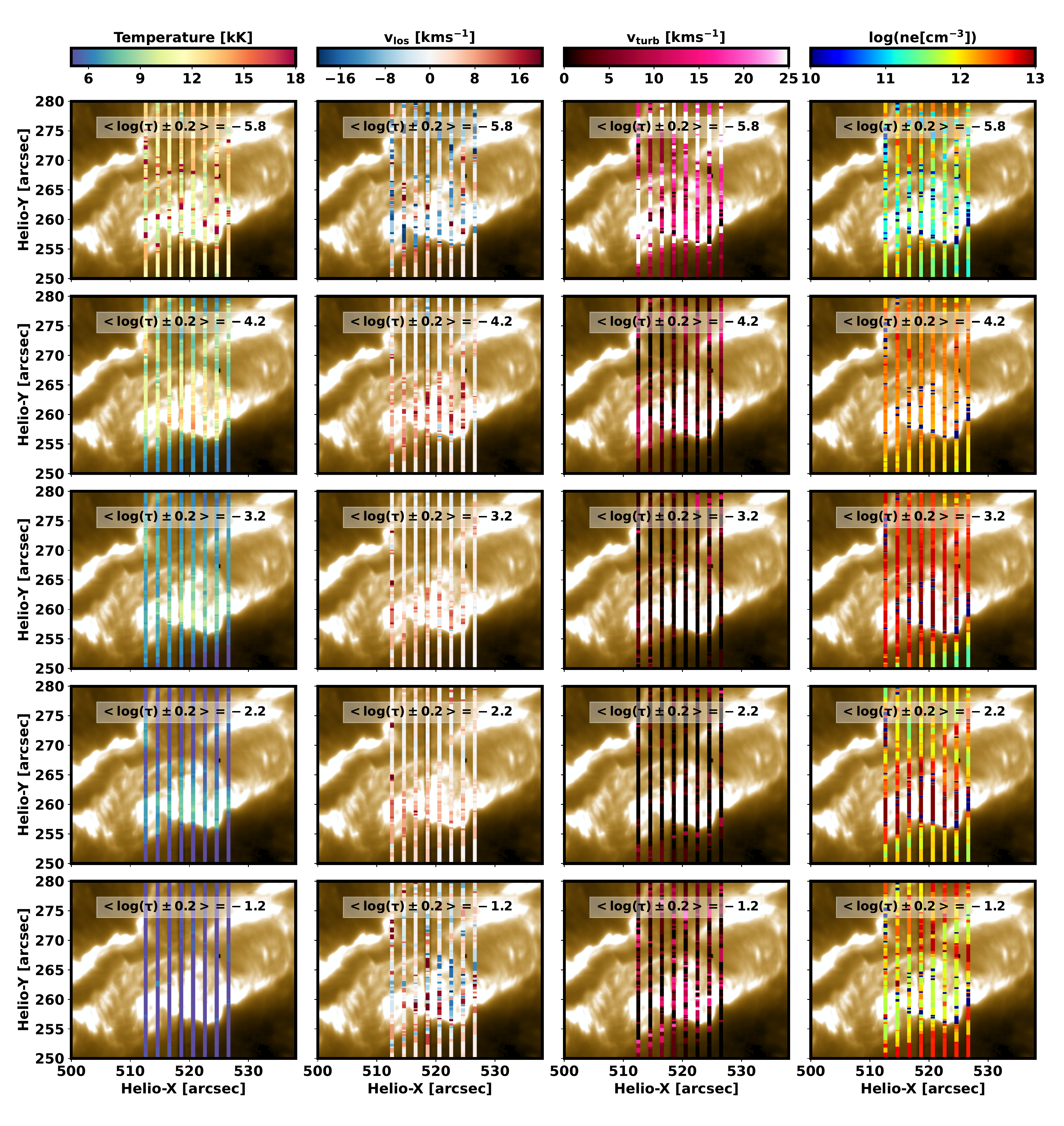}
	\end{center}
	\caption{Maps of thermodynamic values (columns) during the maximum of the X1.0-class flare SOL2014-03-29T17:48, for various optical depths (rows).}\label{fig:allinv}
\end{figure*}

The inversion and interpretation of the profiles presented in this paper entail a significant challenge. The highly dynamic event studied - the maximum of an X1.0-class flare, is reflected both in the associated profiles and the thermodynamics recovered from the inversions of these profiles. The extremely pointy and combined type profiles belong to the same solar feature - the ribbon, including its leading and trailing edge. However, they are related to slightly different stages of the same event, which happen simultaneously in different location in the ribbon. We should understand that both the variation in the appearance of the profiles and their associated thermodynamics is gradual. Thus, while we have focused our attention on the most significant of each type, the following interpretation capture the main physical properties during the maximum of the flare. Figure \ref{fig:allinv} is particularly helpful for this interpretation.

For the positions scanned by the IRIS slit, the location of the trailing edge of the upper ribbon is at $[X,Y] = [512,268]$ and in the lower ribbon at $[X,Y] = [517-526,263]$. It is in these locations where the extremely pointy type A profiles such as the ones shown in Fig. \ref{fig:typeA} are found, while the rest of the profiles of this type are mostly located within the ribbon itself. The location for the leading edge of the upper ribbon is at $[X,Y] = [512,277]$ and in the lower ribbon at  
$[X,Y] = [512-526,257-255]$. Most of the  type B pointy profiles are located at the leading edge of the lower ribbon, while the combination profiles are located within the ribbon on the trailing side immediately adjacent to the leading edge and the ribbon of the ribbon (some are also located in the region just leading the trailing edge of the ribbon). Thus, as the ribbon is energized, starting from the trailing edge towards the leading edge, the profiles go from extremely pointy type A to the combination type and finally to the type B.  

Because ribbons propagate across the solar surface as the flare evolves, the spatial distribution of the thermodynamic parameters from trailing to leading edge of the ribbon provide a window into the typical temporal evolution within a single location. Is is interesting to note that the trailing edge has been energized longer than the leading edge in this single snapshot. 
That could explain why in these locations the temperature is so high in the high chromosphere (\ltau$ = -5.8$, first panel in the first row of Fig. \ref{fig:allinv}), while in the rest of the ribbon the high chromosphere temperature is lower. The trailing edge also differs from the rest of the ribbon when we consider the temperature difference between the high and middle chromosphere. In the trailing edge the temperature of the high chromosphere is higher than in the mid chromosphere. In contrast, the mid chromosphere temperature (\ltau$ = -4.2$, first panel of the second row of Fig. \ref{fig:allinv}) is higher than in the high chromosphere for the leading edge and interior of the ribbon itself. Thus, the bump in the temperature at \ltau$\approx-5$ observed in Fig. \ref{fig:inv_typeB} has not reached yet the high-chromosphere, as it does in Fig.  \ref{fig:inv_typeA_2} and even higher in Fig.  \ref{fig:inv_typeA_1}. This spatial pattern of a somewhat reduced temperature in the high chromosphere accompanied by an increase in the mid chromosphere is seen in both flare ribbons but most clear in the lower ribbon as IRIS scanned this ribbon more fully. This pattern has been obtained in some numerical models of flare energy deposition in the chromosphere by \citealt{Allred15}.  These observations support a scenario in which energy is deposited in the middle chromosphere, with associated local temperature increase. This energy deposition affects the high chromosphere at later times, as the flare evolves. 


The most critical physical parameter that contributes to the very distinctive profiles studied in this paper is the line-of-sight velocity. As we have demonstrated, the extremely pointy profiles of the \mgii\ lines, but also of the \cii\ lines, need the presence of a strong, divergent velocity gradient located between the high and middle chromosphere. In addition, the \mguv\ lines have signatures associated with strong velocities in the high photosphere. The divergent flow located between the high- and mid-chromosphere can be appreciated between the first (i.e., top) and the second panel of the second column of Fig. \ref{fig:allinv}. There, we can observe predominantly an upflow in the ribbon in the high chromosphere (\ltau$ = -5.8$), while in the middle chromosphere and lower regions in the atmosphere ($-4.2 <$\ltau) the ribbon shows a downflow. Note that in the trailing edge, there are some locations where the velocities in the high chromosphere are positive, i.e., they host downflows. Again, these locations are likely {\it ahead in time} (i.e., have evolved for the longest time since the start of the flare), so it is possible they may have experienced the upflows at an earlier stage of their evolution. The presence of a divergent flow is compatible with an scenario where an electron beam propagating downwards from the flare reconnection site in the corona impacts the dense chromosphere (thick-target model, \citealt{Hudson72a}). Such divergent flows have also been obtained in radiation hydrodynamic experiments by \citealt{Kerr16} and \citealt{Kowalski17} who studied the same flare that we analyzed in the current paper. However, the synthetic \mgii\ profiles obtained by \citealt{Kerr16} show the $k_3$ feature in absorption, which indicates that their models are missing some ingredient(s) needed to reproduce the observed profiles. Similarly, the temperature increase where the divergent flows occur in \citealt{Kowalski17} is much higher ($T\approx 10MK$) than the one we obtain. 
What can explain the strong velocity flows that we observe in the high photosphere? Several results suggest that these can be explained by the different penetration of the different energy regimes of the electron beams. \citealt{Graham20} demonstrated that low-energy electrons ($E\approx25-50~keV$) are  responsible for the evaporation-condensation in the high chromosphere, while very high-energy electrons ($E\ge~50~keV$) can penetrate deeper in the atmosphere and produce a similar situation in the high photosphere. These authors used, in addition to the \mguvo\ line, the optically thin lines \fei\ 2814.11 \AA, and \feii\ 2813.3 and 2814.45 \AA. In their study all these lines show a strong red component. However, in the flare we study here no red components are present in the \fei\ and \feii\ lines, and the \mguvo\ line shows on occasions a strong red component, but also a blue and red components in other cases. Likely, the situation shown in Fig. \ref{fig:allinv} is compatible with the scenario described by \citealt{Graham20} and the one previously suggested by \citealt{Libbrecht19} (see Fig. 14 of their paper). Here, in agreement with \citealt{Graham20}, we interpret that the  bounce back motion that \citealt{Libbrecht19} locates in the chromosphere is lower in our case, reaching the high-photosphere. In summary, there are two slabs, one located between the high- and mid-chromosphere and the other in the high photosphere, that are suffering the impact from energized electron coming from the corona, and producing  explosive, divergent upflows and downflows. These slabs are not necessary located in the same feature of the ribbon at the same time, and their location evolves as the ribbon is energized by the flare.

Finally, the turbulent motions are rather strong in the high and middle chromosphere in the ribbons. It is in these regions where the core of both the \cii\ lines and the \mgii\ lines is sensitive to this parameter. Therefore these values are rather feasible. In the low chromosphere and the high photosphere the \vturb\ in the ribbons is negligible.  The extended, broad wings of the \mgii\  are sensitive to changes in \vturb\ in the mid chromosphere, while the \mguv\ lines are sensitive to turbulence in the low chromosphere. We also note that the electron density in the ribbons is $12<\log(n_e)<13$ in the chromosphere, reaches its maximum in the low chromosphere with a value of  $\log(n_e)\approx 15$, just before the temperature minimum, which is {\it pushed down} (in log $tau$ terms) towards the low-photosphere (\ltau$\approx-1$).

The spectral profiles studied in this article are challenging to model and interpret due to the complexity of the physical conditions that generates them. In addition to belonging to an extreme event (an X1-class flare), we note that several physical processes such as upflows and downflows, or heating and cooling, occur simultaneously in the same structure - the ribbon. We are analyzing the maximum of the flare, but that does not mean all regions in the ribbon show the same behavior. For example, the trailing edge experiences different physical conditions than the ribbon itself. The interpretation of the peculiar profiles clearly will depend on where and when they are observed in the macroscopic spatio-temporal evolution of the flare. 
Until now there have not been any theoretical or numerical models that are able to properly reproduce these profiles. In this context, our investigation and results provide strict observational constraints to these models, and suggest a reasonable physical scenario. 

\section*{Conflict of Interest Statement}

The authors declare that the research was conducted in the absence of any commercial or financial relationships that could be construed as a potential conflict of interest.

\section*{Author Contributions}

This paper has been led by ASD. He has made the analysis of the IRIS spectra, and the interpretation of the inversions and the models recovered from them, as well as the main contain of the manuscript . BDP has contributed with meaningful ideas, enlightening discussions with ASD, and with critical comments and improvements to the manuscript. Both authors have reviewed this manuscript and approved it for publication.  

\section*{Funding}
ASD was supported in part by the NASA contract NNG09FA40C (IRIS) and the NASA grant 80NSSC21K0726 {\it ``Understanding the role played by the
turbulence in the chromosphere during flares''}. BDP was supported by NASA contract NNG09FA40C (IRIS). 

\section*{Acknowledgments}
IRIS is a NASA small explorer mission developed and operated by LMSAL with mission operations executed at NASA Ames Research center and major contributions to downlink communications funded by ESA and the Norwegian Space Agency. The authors thank the helpful discussions and clarifications made by J. de la Cruz Rodríguez. We also thanks him for sharing the state-of-the-art inversion code STiC, which is available to the public.


\section*{Data Availability Statement}
The original datasets used in  this investigations are available in a publicly accessible repository. They can be found \href{https://www.lmsal.com/hek/hcr?cmd=view-event&event-id=ivo%3A%2F%2Fsot.lmsal.com%2FVOEvent%23VOEvent_IRIS_20140329_140938_3860258481_2014-03-29T14%3A09%3A382014-03-29T14%3A09%3A38.xml}{here}, or 
through the IRIS L2 data search tool (\href{https://iris.lmsal.com/search/}{https://iris.lmsal.com/search/}).

\bibliographystyle{Frontiers-Harvard} 
\bibliography{allbib,others}




\newpage

\newpage

\end{document}